\documentclass[prl,superscriptaddress,showpacs,amssymb,amsmath,
amsfonts,aps,twocolumn,secnumarabic,floatfix]{revtex4}
\usepackage{graphics}
\usepackage{epsfig}
\begin{document} 
\bibliographystyle{try} 
 
\topmargin 0.1cm 
 
 \title{The electro-disintegration of few body systems revisited}

\newcommand*{\SACLAY }{ CEA-Saclay, Service de Physique Nucl\'eaire, F91191 Gif-sur-Yvette, Cedex, France} 
\affiliation{\SACLAY } 

\newcommand*{\JLAB }{ Thomas Jefferson National Accelerator Facility, Newport News, Virginia 23606} 
\affiliation{\JLAB } 

\author{J.M.~Laget}
     \affiliation{\SACLAY}
     \affiliation{\JLAB}

\date{\today} 
 
\begin{abstract}

Recent studies of the electro-disintegration of the few body systems at JLab have revived the field. Not only  recoil momentum distributions have been determined in a single shot. But also they confirm that the diagrammatic approach, which I developed 25 years ago, is relevant to analyze them, provided that the Nucleon-Nucleon scattering amplitude, determined in the same energy range, is used. They provide us with a solid starting point to address the issue of the propagation of exotic components of hadrons in nuclear matter  
\end{abstract} 
 
 
\maketitle 

The primary goal of the study of the (e,e$'$p) reaction on nuclei was, and still
is, the determination of the high momentum components of the nuclear wave
function. In the past, the spectral functions measured at Saclay or Amsterdam suffered from large corrections (about a factor two or more) due to Final State Interactions (FSI) and Meson Exchange Currents (MEC). A survey of the  state of the art at that time can be found in ref.~\cite{La91}. The corresponding experiments were performed at low values ($\sim 0.4$~GeV$^2$) of the virtuality $Q^2$ of the exchanged photon.

When it was decided to build CEBAF, a common belief was that increasing $Q^2$ was the way to suppress FSI and MEC contributions. This is partly true, since both the FSI and MEC amplitudes involve a loop integral, which connects the nuclear bound and scattering states and which is expected  to decrease when $Q^2$ increases as form factors do. But this is partly wrong, since the singular part of the FSI integral does not depend on $Q^2$, besides the trivial momentum dependency of the elementary operators. It comes from unitarity, and corresponds to the propagation of an on-shell nucleon. It involves on-shell elementary matrix elements and it is maximum when the kinematics allows for rescattering on a nucleon at rest~\cite{La81}. In the (e,e$'$p) channel, this happens in quasi-free kinematics, when $X=Q^2/2m\nu=1$ ($\nu$ being the energy of the virtual photon, and $m$ the nucleon mass).

In turn, this kinematics provides us with a way to isolate NN scattering (or more generally scattering between hadrons) and opens up an original use of the (e,e$'$p) reactions~\cite{La98,La00}: the study of exotic components of the hadron wave function via  color transparency or color screening, for instance. 

\begin{figure}[hbt]
\begin{center}
\epsfig{file=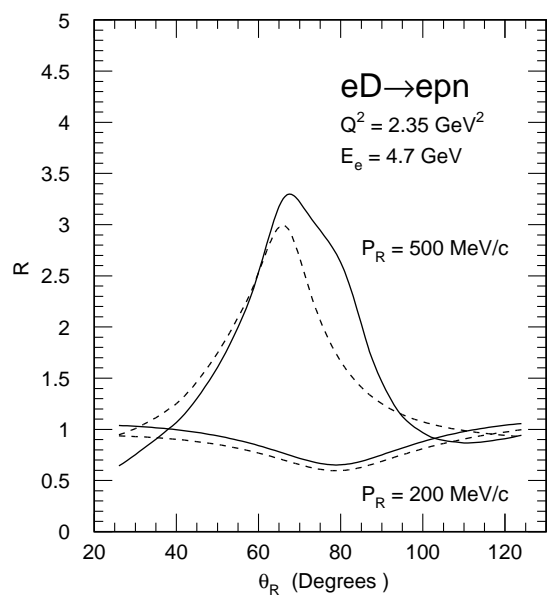,width=3.0in}
\caption[]{The ratio between the full cross section and the contribution of the
quasi-free scattering.}
\label{sing}
\end{center}
\end{figure}

Fig.~\ref{sing} exhibits these features. It shows the angular distribution,
against the neutron angle $\theta_R$ with the virtual photon, of the ratio between the full cross section of the D(e,e$'$p)n reaction and the quasi-free contribution, when the momentum $P_R$ of the recoiling neutron is kept constant. FSI (dashed curves) are maximum near $\theta_R=70^{\circ}$ where $X=1$ and on-shell rescattering is maximized. At low values of the recoil momentum ($P_R= 200$~MeV/c), on-shell nucleon rescattering reduces the quasi-free contribution, as expected from unitarity (a part of the strength of the quasi-elastic channel is transferred to inelastic ones).  At high values of the recoil momentum ($P_R= 500$~MeV/c) the quasi-free contribution strongly decreases as the nucleon momentum distribution: on-shell rescattering takes over and dominates. 

Similarly, the $\Delta$, which is produced on one nucleon and exchanges a meson with the second nucleon in the MEC amplitude, can also propagate on-shell. The corresponding singularity appears at larger recoil angles and shifts the NN rescattering peak (full curves). In fact other baryonic resonances can be excited and propagate, widening the peak further toward larger angles. But the $\Delta$ is the most prominent part of the nucleon response function, and the effects of the higher mass resonances are expected to be smaller, except maybe at higher recoil momenta.

Experiments~\cite{Bo03,Kim04} recently performed at JLab confirm this behavior, which was already predicted~\cite{La78} and measured~\cite{Ar78} in the $\pi$N rescattering sector at lower energy.

To be more specific, the method~\cite{La81} is based on the expansion of the amplitude in terms of few relevant diagrams, which are computed in the momentum space, in the Lab. frame. The kinematics as well as the propagators are relativistic and no angular approximation is made in the evaluation of the loop integrals. The elementary operators which appear at each vertex have been calibrated against the corresponding channels. Its application to the D(e,e$'$p)n channel has been discussed in refs~\cite{La78b,La84} and to the $^3$He(e,e$'$p) channels in refs.~\cite{La85,La87}. A comprehensive summary is given in ref.~\cite{La94} for the $^4$He(e,e$'$p)T channel. 

\begin{figure}[hbt]
\begin{center}
\epsfig{file=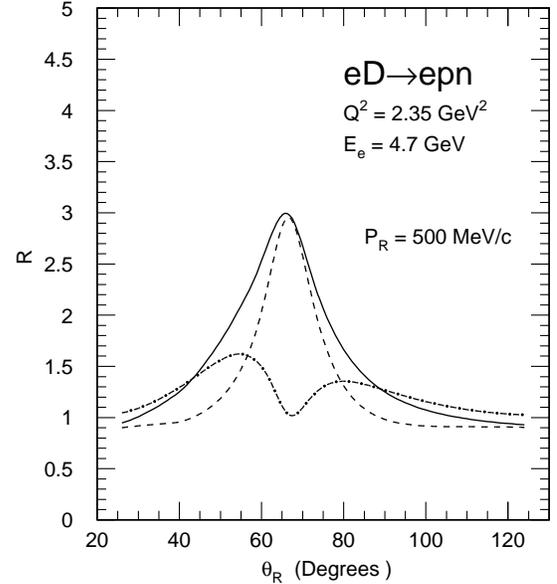,width=3.0in}
\caption[]{The angular distribution of the on-shell (dot-dashed) and off-shell (dashed) parts of the ratio of the FSI to the PW cross sections.}
\label{on-off}
\end{center}
\end{figure}

For the sake of the discussion, I reproduce the Plane Wave (PW) and FSI amplitudes for the D(e,e$'$p)n channel
\begin{eqnarray}
T_{PW}= \sum_{m_p} \langle m_1{\mid}J_p(q^2){\mid}m_p\rangle
 \langle \frac{1}{2}\, m_p\, \frac{1}{2}\, m_2{\mid}1M_{J}\rangle
 U_0\left( \vec{p}_2\right) \frac{1}{\sqrt{4 \pi}} \nonumber \\
  +\sum_{m_n} \langle m_2{\mid}J_n(q^2){\mid}m_n\rangle
 \langle \frac{1}{2}\, m_n\, \frac{1}{2}\, m_1{\mid}1M_{J}\rangle
 U_0\left( \vec{p}_1\right) \frac{1}{\sqrt{4 \pi}} \nonumber \\
 + \mathrm{D}\; \mathrm{ Wave} \;\;
\end{eqnarray} 
\begin{eqnarray}
T_{FSI}=\sum_{\lambda_p\lambda_nm_lm_s}\int\frac{d^3\vec{n}}{(2\pi)^3}
\frac{m}{E_p(p^0-E_p+i\epsilon)}
 \nonumber\\ 
\left\{(\lambda_p\mid J_p(q^2)\mid m_s-\lambda_n)
(\vec{p}_1m_1\vec{p}_2m_2\mid T_{NN}\mid \vec{p}\lambda_p\vec{n}\lambda_n)
\right. \nonumber \\ 
\left. +(\lambda_p\mid J_n(q^2)\mid m_s-\lambda_n)
(\vec{p}_2m_2\vec{p}_1m_1\mid T_{NN} \mid \vec{p}\lambda_p\vec{n}\lambda_n)\right\}
\nonumber \\ \times (\frac{1}{2}\lambda_n\frac{1}{2}(m_s-\lambda_n)\mid 1m_s)
\left\{\frac{1}{\sqrt{4\pi}}U_0(\mid\vec{n}\mid)
\delta_{M_Jm_s}
\delta_{m_l0}\right.\nonumber\\
\left.+U_2(\mid\vec{n}\mid)
(2m_l1m_s\mid 1M_J)
Y_2^{m_l}(\widehat{\vec{n}})
\right\} 
\label{fsi}
\end{eqnarray}
where $E_p=\sqrt{m^2 +(\vec{k}-\vec{n})^2}$ and  $p^0=M_D + \nu -\sqrt{m^2+\vec{n}^2}$.
The momenta and magnetic quantum numbers of the outgoing proton and neutron are respectively $\vec{p}_1$, $\vec{p}_2$, $m_1$ and  $m_2$, while the magnetic quantum number of the target deuteron is $M_J$. The $S$ and $D$ parts of the deuteron wave function are respectively $U_0$ and $U_2$. The relativistic expressions of the proton $J_p(q^2)$ and neutron $J_n(q^2)$ currents are used in both the PW and FSI amplitudes, contrary to~\cite{La94} where their expansion up to and including terms of order $1/m^3$ was used: the difference does not exceed a few per cent, except at very forward or backward recoil angles. The FSI integral runs over the momentum $\vec{n}$ of the spectator nucleon. Since the energy is larger than the sum of the masses of the two nucleons, the knocked out nucleon ($\vec{p},\lambda_p$) can propagate on-shell. Due to the dominance of the $S$-wave part of the wave function, the corresponding singular part of the integral is maximum when  the scattering of the electron on a nucleon at rest is kinematically possible (see ref.~\cite{La81} for a full discussion): This happens in the quasi-elastic kinematics, $X=1$. The width of the on-shell peak in~Fig.~\ref{on-off} reflects the Fermi distribution of the target nucleon, while the off-shell (principal) part of the integral vanishes at $X=1$.

The physical picture is the following. The electron scatters on a proton at rest which propagates on-shell and rescatters on the neutron which is also at rest. In the Lab. frame, the soft neutron recoils at 90$^{\circ}$ with respect to the fast proton which is emitted in the forward direction. Two body kinematics imposes that the angle of the rescattering peak (dip) moves  with the recoil neutron momentum: around 70$^{\circ}$ when $p_R=$ 500~MeV/c,  80$^{\circ}$ when $p_R=$ 200~MeV/c. The same occurs, in a different part of the phase space, when the electron interacts with the neutron.  In the classical Glauber approximation, the nucleon propagator in Eq.~\ref{fsi} is linearized and recoil effects are neglected: Therefore the rescattering peak stays at 90$^{\circ}$~\cite{Je96,Cio01}. This drawback has been cured in the Generalized Eikonal Approximation (GEA)~\cite{SaXX} which takes into account higher order recoil terms in the nucleon propagator, and neglects only terms of the order $p_{\perp}^2/m^2$. It comes as no surprise that GEA predicts the FSI peak at the same place as in my diagrammatic approach which takes into account the full kinematics from the beginning~\cite{La81,La94}. While it is valid at forward angles, the classical Glauber treatment is simply not correct for analyzing the (e,e$'$p) reactions at large angles and large recoil momentum.

Since it involves on shell matrix elements and relies on the low momentum components of the wave function, the FSI amplitude is founded on solid ground near $X=1$, provided the correct parameterization of the NN amplitude is used.

In the pre-CEBAF era, the relative kinetic energy of the two outgoing nucleons ($T_L= Q^2/2m\simeq 200$~MeV) was low enough to rely on the partial wave expansion of the nucleon-nucleon  scattering amplitude($T_{NN}$), see for instance~\cite{La87,La94}, of which both the on-shell and half off-shell parts were solutions of the Lippman-Schwinger equation with the same potential (Paris) as for the bound state~\cite{PaXX}. S, P and D waves were retained and the FSI loop integral was done analytically according to ref.~\cite{La87a}, fully taking into account Fermi motion effects (unfactorized calculation). When this is done, and the momenta expressed in the rest frame of the neutron-proton system,  Eq.~\ref{fsi} coincides with Eq.~C.8 of~\cite{La94}. 

\begin{figure}[hbt]
\begin{center}
\epsfig{file= 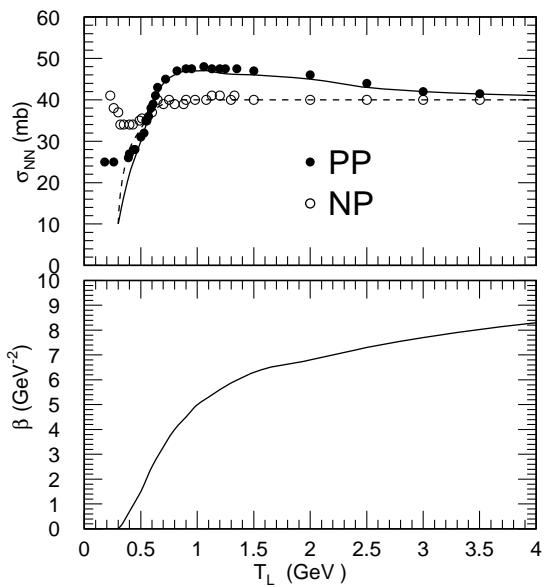,width=3.0in}
\caption[]{The variation of $\sigma_{NN}$ and $\beta$ with energy.}
\label{beta}
\end{center}
\end{figure}

At higher energies (let's say when the relative kinetic energy of the outgoing
fragments exceeds 500~MeV or so), too many partial waves enter into the game and
their growing inelasticities prevent us to compute the scattering amplitude from a
potential. It is better to use a global parameterization of the NN scattering
amplitude. On general grounds~\cite{Bi78,Mc83}, it can be expanded as follows
\begin{equation}
T_{NN} = \alpha + i\gamma (\vec{\sigma_1}+\vec{\sigma_2})\cdot \vec{k}_{\perp}
+\;\mathrm{spin-spin} \;\mathrm{terms} 
\label{NN}
\end{equation}
where $\vec{k}_{\perp}$ is the unit vector perpendicular to the scattering plane.

Above 500 MeV, the central part $\alpha$ dominates. It is almost entirely
absorptive, and takes the simple form
\begin{equation}
\alpha = -\frac{Wp_{cm}}{2m^2} \;(\epsilon + i)\;\sigma_{NN} \;\exp[\frac{\beta}{2}t] 
\end{equation}
In the forward direction its imaginary part is related to the total
cross section $\sigma_{NN}$, while the slope parameter $\beta$ is related to the
angular distribution of NN scattering. Both can be determined from the
experiments performed at Los Alamos, Saturne and COSY. Fig.~\ref{beta} shows the values which I use. Below 500 MeV, I have extrapolated them in such a way the absorptive part of the amplitude vanishes at the pion production threshold. The ratio $\epsilon$, between the real and imaginary part of the amplitude, is small: I keep it constant ($\epsilon=-0.2$) above 1 GeV, and smoothly extrapolate it down to zero at the pion threshold. 

Such a parameterization is very  convenient to compute the rescattering 
amplitude. It adds its absorptive part, which dominates at high energy, to its expansion in terms of the real part of phase shifts (of which  I use the experimental values, above $T_L= 500$ MeV), which dominates at low energy. However, at high energies, it leads only to an accurate prediction of its singular part (on-shell scattering). Contrary to low energy, there is unfortunately  no way to constrain the half-off shell behavior of the absorptive  part of the NN scattering amplitude, and one can get only an estimate of the principal part of the rescattering amplitude. It turns out that it vanishes at X=1 (Fig.~\ref{on-off}) and it does not dominate at high energy. So, the method is founded on solid grounds in the quasi-elastic kinematics (X$\sim$1). Away, it tells us in which kinematics FSI are minimized. 

\begin{figure}[hbt]
\begin{center}
\epsfig{file=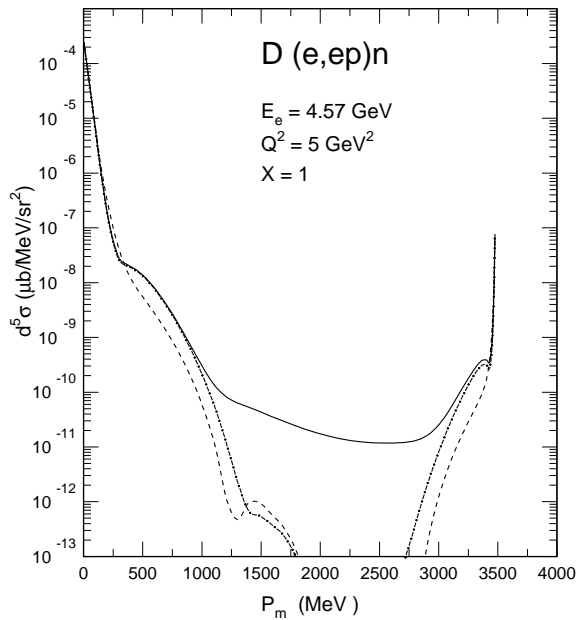,width=3.0in}
\caption[]{The momentum distribution in the D(e,e$'$p)n reaction at X=1 and
$Q^2=5$~GeV$^2$. Dashed line: PW. Dash-dotted line: with FSI. Full line: MEC and $\Delta$ included.}
\label{d_x1q5}
\end{center}
\end{figure}

The  $\Delta$ formation and MEC amplitudes are computed according to eqs~C1 and C2 in ref~\cite{La94}. I updated them by implementing the full relativistic expression of the $\pi NN$ vertex and using the latest Q$^2$ dependency of the $N\rightarrow \Delta$ electromagnetic form factor~\cite{DeEM}. Since it falls down more rapidly than the Nucleon form factor, the $\Delta$ formation amplitude is suppressed at high Q$^2$. Also the unitary singularity associated with the $\Delta$ propagation is weaker than in the FSI amplitude since the  $\Delta$ pole is distant from the energy axis by its half width. Again the $\Delta$ propagates almost on shell in the kinematics of Fig.~\ref{sing}, and it is worth to emphasize that the parameters are those which reproduce the $NN\rightarrow N\Delta$ cross section in the few GeV range (see e.g.~\cite{La93}).
 
Fig.~\ref{d_x1q5} shows the full angular distribution of the D(e,e$'$p)n reaction for $Q^2= 5$~GeV$^2$, at the top of the unitary peak in Fig.\ref{sing}, $X=1$.  The $\Delta$ formation term contributes little up to $p_n \sim 800$~MeV/c, but dominates above. At the extreme backward proton emission angles (large momentum of the neutron but vanishing momentum of the proton) the interaction of the electron with the neutron takes over and is modified, as the forward proton peak, by FSI, MEC and $\Delta$ formation term. These findings are reproduced by the preliminary analysis of the D(e,e$'$p)n reaction~\cite{Kim04} recently recorded in the full phase space with CLAS at JLab. We must await its final analysis for a detailed comparison.

\begin{figure}[hbt]
\begin{center}
\epsfig{file=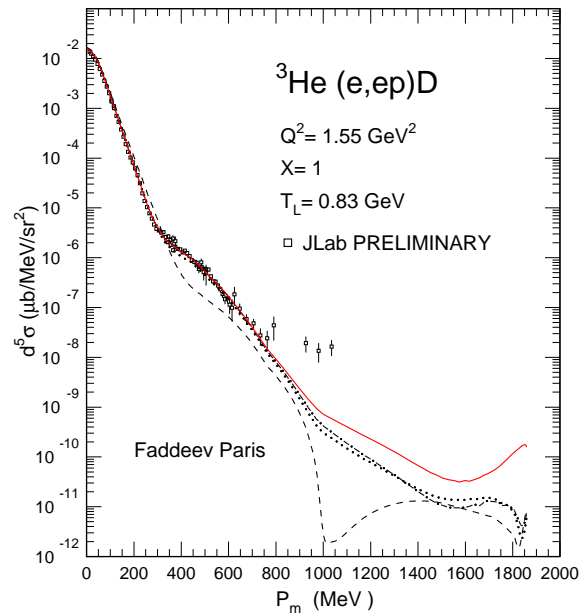,width=3.0in}
\caption[]{The momentum distribution in the $^3$He(e,e$'$p)d reaction at X=1 and
$Q^2=1.55$~GeV$^2$. Dashed line: PW. Dotted line: with FSI. Dash-dotted line: 2 body MEC and $\Delta$ included. Full line: 3 body mechanisms included.}
\label{he3pd}
\end{center}
\end{figure}

So far, only the analysis of the electro-disintegration of $^3$He and $^4$He have been completed at JLab. Fig.~\ref{he3pd} shows how well the diagrammatic method reproduces the cross section of the
$^3$He(e,e$'$p)d reaction recently measured~\cite{Hig02,Ma03} with two magnetic spectrometers,  at $Q^2=1.55$~GeV$^2$, in the quasi-free kinematics (X=1). The wave function is the solution~\cite{Sau81,Sau83} of the Faddeev equations for the Paris potential~\cite{PaXX}. The nucleon single scattering (FSI) and two body MEC amplitudes are implemented as described in ~\cite{La94}. Both $pp$ as well as $T=0$ and $T=1$ $np$ active pairs are considered. At such a high virtuality, the relative kinetic energy between the outgoing proton and deuteron is $T_L= 830$~MeV, where the NN cross section reaches its maximum and becomes flat around $\sigma_{NN}= 45$~mb. Again, FSI reduces the quasi-free contribution below 300~MeV/c and overwhelms it by more than a factor five around 500~MeV/c. Above 1~GeV/c, MEC and $\Delta$ production enhance the cross section, but are unable to reproduce the last three experimental points around 1~GeV/c. Here, one enters into the kinematical regime where the deuteron is fast and emitted in the forward direction while the proton is slow and becomes a spectator: this is responsible for the small deuteron knockout peak at the extreme right of the figure. In order to accommodate the experiment around 1~GeV/c and above, one needs a mechanism which shares the photon momentum between the three nucleons. Three body meson rescatterings, computed as in ref.~\cite{La88}, go in the right direction but fall short. It is very likely that nucleon double scattering will finish the job: It provides a way to share the momentum transfer in such a way that a slow proton recoils while two fast nucleons are  emitted in the forward direction, with a small enough relative momentum to recombine into the deuteron. This study remains to be done.

\begin{figure}[hbt]
\begin{center}
\epsfig{file=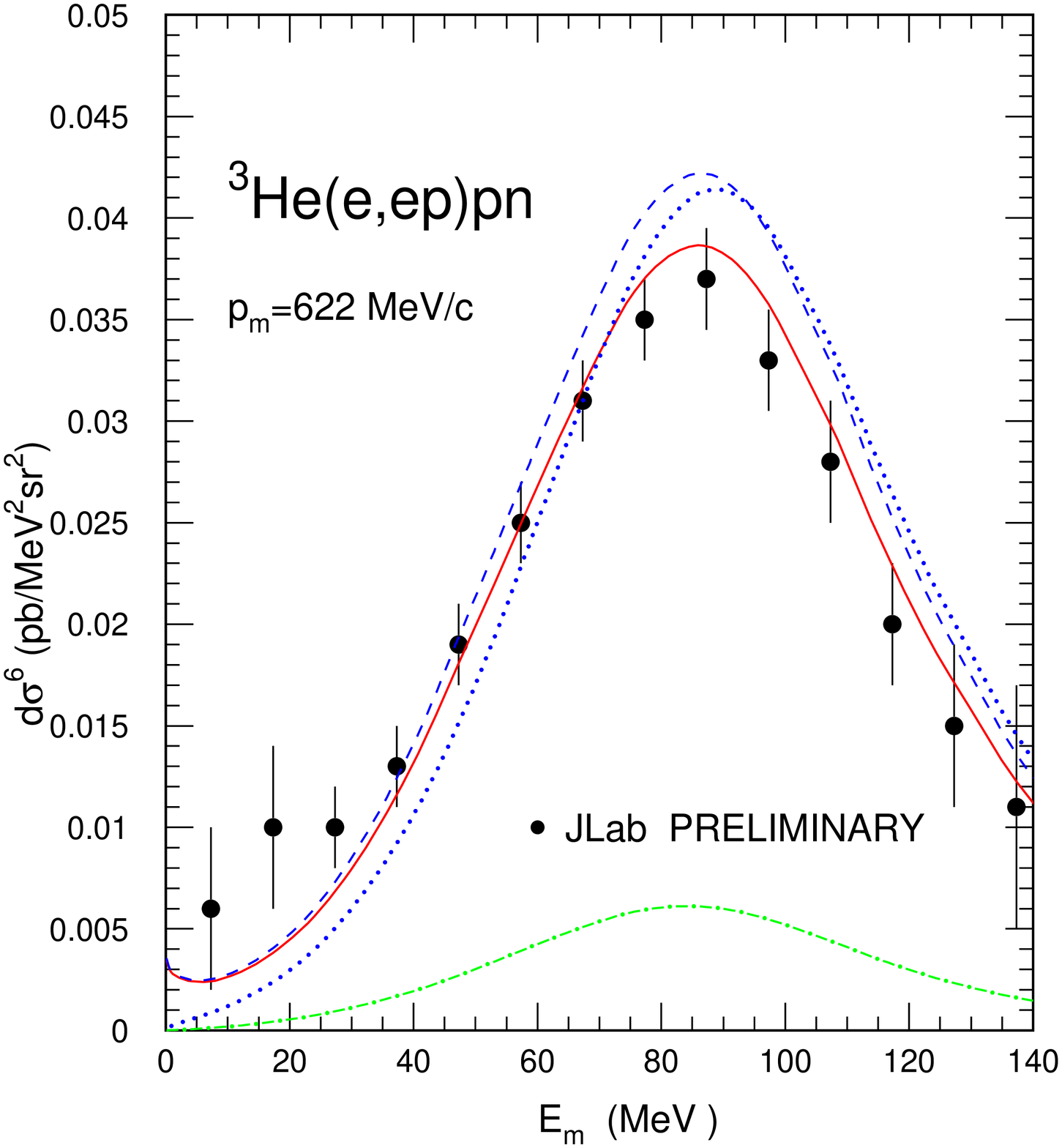,width=3.in}
\caption[]{The recoil energy distribution in the $^3$He(e,e$'$p)np reaction at $p_m$= 622~MeV/c, X=1 and $Q^2=1.55$~GeV$^2$.}
\label{cont}
\end{center}
\end{figure}

In the same experiment, the $np$ continuum has been recorded. 
Two body short range correlations are the primary source of high momentum
components in the nuclear wave function. They are strongly coupled to high
energy states in the continuum, where they induce a peak (dot-dashed line in Fig.~\ref{cont}) characteristic of the disintegration of a NN pair at rest in $^3$He~\cite{La91}. The width of the peak reflects the Fermi motion of the pair. Again, FSI between the two nucleons of the pair (dotted line) dominate the cross section. The subsequent scattering of one of these nucleons with the spectator third nucleon (dashed line) shifts the peak toward the experiment~\cite{FaXX}, but is not dominant. MEC and $\Delta$ formation (full line) brings down the cross section in good agreement with the experiment. In the continuum, one measures the transition between a correlated pair in the $^3$He ground state and a correlated pair in the continuum. It turns out that $pn$ pairs (in $T=0$ and $T=1$ isospin states) as well as $pp$ pairs contribute by roughly the same amount.

Triple coincidence studies~\cite{Zha02,Gi01} of the reaction $^3$He(e,e'NN)N have been completed in the full phase space with the large acceptance spectrometer CLAS at JLab. The model reproduces also the cross section for various cuts in the phase space. I refer to my talk at the Lisboa conference~\cite{La03} for a comparison with preliminary data.

Finally, the model gives a good account of the cross section of the reaction $^4$He(e,e$'$p)T,  which has been recently determined at JLab: see ref.~\cite{Re03} for a comparison with preliminary data.

It is remarkable that the cross sections of so many channels are reproduced with the simple choice (eq.~\ref{NN}) of the central part of the nucleon-nucleon scattering amplitude. While its value at  the very forward angle is fixed by unitarity, the slope parameter $\beta$ has been determined by fitting the angular distribution of the unpolarized nucleon-nucleon scattering cross section. It turns out that this form reproduces fairly well the modulus of the central part of the NN amplitude extracted form the SAID data base~\cite{SAID}  up to four momentum transfer $-t= 0.4$~GeV$^2$, {\it i.e.} $p_m\sim \sqrt{-t}= 0.63$~GeV/c in the quasi-elastic kinematics. However, the SAID ratio $\epsilon$ between the real part and the imaginary part of the scattering amplitude varies from about zero at $p_m=0$ to about one around $p_m=200$~MeV/c and back to zero in the range $300<p_m<700$~MeV/c. Thus, the parameterization~\ref{NN} is very good in the recoil momentum range where FSI dominate. Above, $\Delta$ formation and MEC take over, and the details of the NN amplitude are less important. The full implementation of the actual SAID amplitudes poses the problem of their extrapolation in the unphysical region (off-shell NN scattering) which is under study. 

\begin{figure}[hbt]
\begin{center}
\epsfig{file=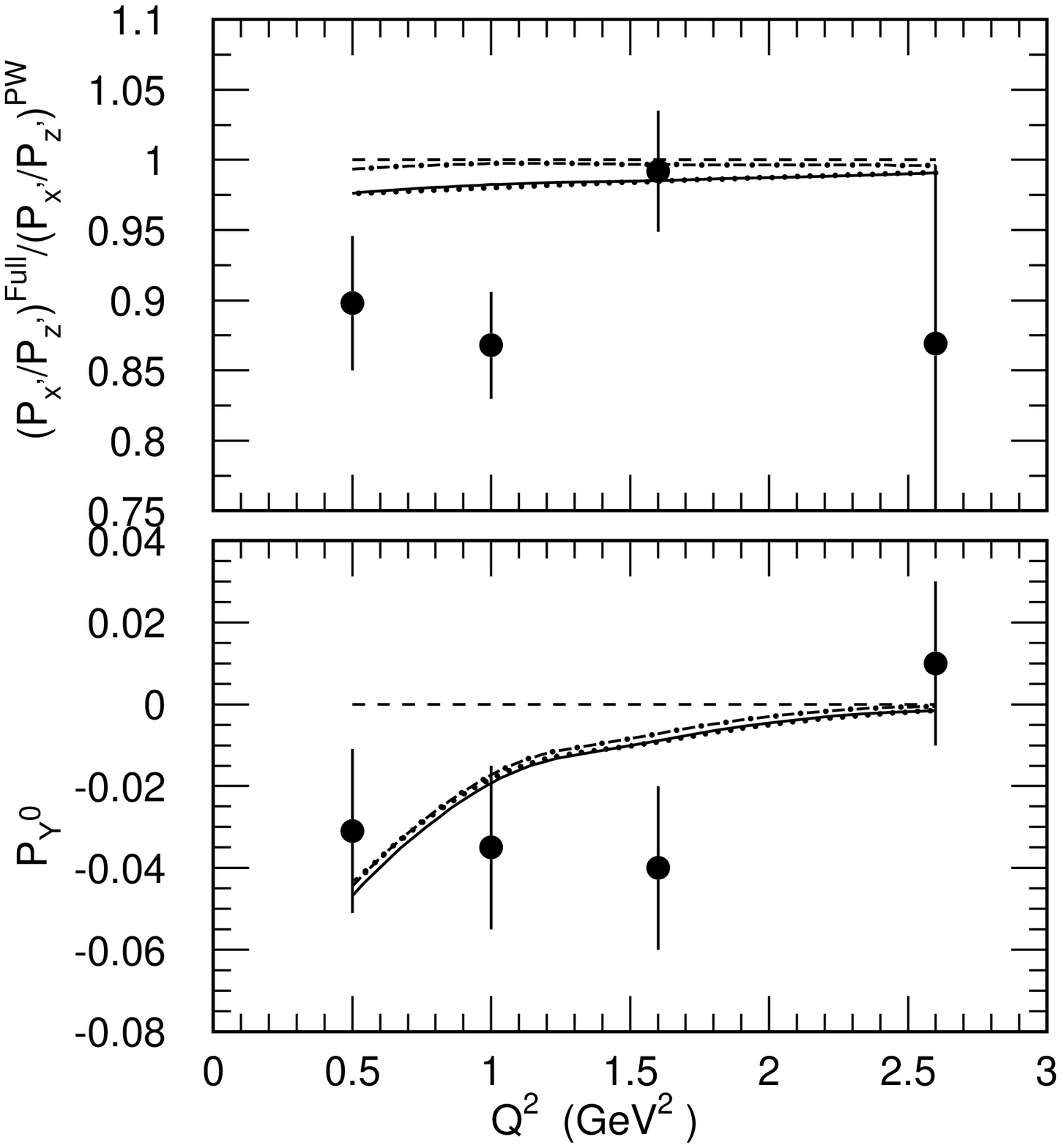,width=3.in}
\caption[]{The ratio of the spin transfer coefficients (top) and the induced proton polarization (bottom) in $^4$He(e,e$'$p)T. Dashed lines: PW. Dash-dotted lines: FSI. Dotted lines: 2 body MEC and $\Delta$ included. Full lines: 3 body mechanisms included.}
\label{spin}
\end{center}
\end{figure}

While the spin-orbit and spin-spin terms are taken into account in the phase shift expansion (low energy) they have not yet been implemented  in the absorptive amplitude (high energy). The spin-orbit term has been found to give small contribution to unpolarized observables in the (e,e$'$p) channels~\cite{Je99}. The present version of the model  predicts deviations from PW of the ratio of spin transfer coefficients (as defined in Appendix A of ~\cite{La94}) of a few~\% in the direction of experimental values~\cite{StXX} recently recorded at JLab (Fig.~\ref{spin}). The good agreement with the induced polarization $P^0_Y$ gives confidence on the treatment of various interaction effects. However, before drawing any definite conclusion, one has to wait until the full implementation of spin dependent terms as well as the averaging over the experimental acceptance.

To summarize, a fair agreement with the recent JLab data has been reached around $X=1$, up to recoil momentum of the order of 1 GeV/c, provided
that the NN scattering amplitude relevant to the same energy range as well as realistic few body wave functions are used. The perpendicular kinematics offers a robust starting point to study the evolution with $Q^2$ of the re-interaction of nucleons, but also of hadrons, in view of determining their structure at short distances~\cite{La98,La00}. It is not the right place to determine the high momentum components of the nuclear wave function. One has to go away from the quasi-elastic kinematics: as demonstrated in Fig.~\ref{sing}, this occurs  in parallel or anti-parallel kinematics, where on-shell nucleon rescattering is suppressed.

I acknowledge the warm hospitality of Jefferson Laboratory, where this work was completed, as well as the numerous discussions which I had with D. Higinbotham, K. Egiyan and E. Voutier over the past few years and which have shaped this project.

\end{document}